\documentclass[aps,twocolumn,showpacs,preprintnumbers,superscriptaddress]{revtex4-2}

\usepackage{graphicx}  

\usepackage[format=plain,font=small,labelfont=bf,justification=centerlast,singlelinecheck=false]{caption}
\usepackage{subcaption}

\usepackage{parskip}
\usepackage[T1]{fontenc}
\usepackage{dcolumn}   

\usepackage{bm}        
\usepackage{amsfonts}  
\usepackage{amsmath}   
\usepackage{amssymb}   
\usepackage{braket}
\usepackage{natbib}    		
\setcitestyle{square, comma, numbers,sort&compress}

\begin{document}


\title{Telecom Quantum Photonic Interface for a $^{40}$Ca$^+$ Single-Ion Quantum Memory}

\author{Elena Arensk\"otter}
\author{Tobias Bauer}
\author{Stephan Kucera}
\affiliation{Experimentalphysik, Universit\"at des Saarlandes, 66123 Saarbr\"ucken, Germany}
\author{Matthias Bock}
\thanks{} 
\affiliation{Experimentalphysik, Universit\"at des Saarlandes, 66123 Saarbr\"ucken, Germany}
\affiliation{present address: Institut für Quantenoptik und Quanteninformation,
\"Osterreichische Akademie der Wissenschaften, Technikerstraße 21a, 6020 Innsbruck, Austria}
\author{J\"urgen Eschner}
\email{juergen.eschner@physik.uni-saarland.de}
\author{Christoph Becher}
\email{christoph.becher@physik.uni-saarland.de}
\affiliation{Experimentalphysik, Universit\"at des Saarlandes, 66123 Saarbr\"ucken, Germany}

\date{\today}

\begin{abstract}  
Entanglement-based quantum networks require quantum photonic interfaces between stationary quantum memories and photons, enabling entanglement distribution. Here we present such a photonic interface, designed for connecting a $^{40}$Ca$^+$ single-ion quantum memory to the telecom C-band. The interface combines a memory-resonant, cavity-enhanced spontaneous parametric down-conversion (SPDC) photon pair source with bi-directional polarization-conserving quantum frequency conversion (QFC). We demonstrate preservation of high-fidelity entanglement during conversion, fiber transmission over up to 40\,km and back-conversion to the memory wavelength. Even for the longest distance and bi-directional conversion the entanglement fidelity remains larger than 95\% (98\%) without (with) background correction.
\end{abstract}


\maketitle 


\section{Introduction}

Entanglement-based quantum networks are the backbone for many quantum technology applications connecting remote partners, such as distributed quantum computing \cite{Jiang2007}, quantum repeaters \cite{Briegel1998}, quantum key distribution (QKD) with entangled photons \cite{Ekert1991} or networks of quantum sensors \cite{Proctor2018}. Here we consider networks involving matter-based quantum memories communicating via optical photons. To realize such networks, several requirements have to be fulfilled simultaneously: (i) a source of high-rate and high-fidelity entanglement is essential \cite{Simon2007,Wehner2018}; (ii) the spectral characteristics of photonic information carriers and matter-based quantum memories have to match; and (iii) the communication wavelength has to be in a low-loss band of optical fibers. 

Requirement (i) may be addressed by photon pair sources based on spontaneous parametric down-conversion (SPDC) that have shown immense potential as high-fidelity sources of entanglement since their introduction \cite{Kwiat1995}. For their interfacing with atom-based quantum memories, requirement (ii) \cite{Lenhard2015, Specht2011, Rosenfeld2011}, a narrow spectrum of the photon pairs is required. This can be attained by spectral filtering \cite{Brito2016} or by placing the nonlinear crystal inside a cavity that is tuned to be resonant with the desired atomic transition \cite{Luo2015, Tsai2018}. The latter scheme also enables very high pair rates, but it may suffer from reduced state purity due to distinguishability induced by the birefringence of both crystal and resonator. An alternative approach with higher-quality entanglement but with a much lower pair rate is to use single-pass conversion in an interferometric configuration, which eliminates all distinguishability and eliminates the 50\% background of unsplit photon pairs \cite{Fiorentino2004, Kuzucu2008, Kim2006}. Ideally, resonator-enhanced generation and interferometric configuration would be combined. 

For the distribution of entangled photon pairs over long distances in fiber networks, requirement (iii), transmission loss is a crucial issue, limiting the entanglement rate and thereby the secret key rate of QKD. Thus it is advantageous to use wavelengths in the telecom bands (1260\,nm - 1625\,nm), where fiber absorption is minimal. Most of the relevant transitions in atomic or ionic quantum memories, however, are in the visible or near IR regime \cite{Kurz2016, Meyer2015, Schupp2021, Mount2013}. One possible approach to address quantum memory transitions and at the same time minimize fiber loss is to use non-degenerate SPDC entangled-pair sources. In this case one photon is resonant to the memory transition while the other has a wavelength in the telecom regime \cite{Saglamyurek2011, Fekete2013, Bussieres2014, Lenhard2015, Riedmatten2021}. Alternatively, non-degenerate photon pairs may be generated from degenerate pairs by quantum frequency conversion (QFC) of one of the photons \cite{Kumar1990} which shifts the photon wavelength while preserving all other properties. Here, an efficient conversion process is three-wave mixing in $\chi^{(2)}$-nonlinear media \cite{Huang1992}, but its strong polarization dependence is adverse to the conversion of polarization qubits. To overcome this limitation, several schemes for polarization-preserving conversion have been developed \cite{Albota2006, Ramelow2012, Krutyanskiy2017, Bock2018, Ikuta2018, Kaiser2019, vanLeent2020} and employed to demonstrate photon-photon \cite{Ramelow2012, Kaiser2019} and light-matter entanglement over short distances \cite{Bock2018, Ikuta2018} as well as over km-long fibers \cite{Krutyanskiy2019, vanLeent2020, vanLeent2022,Luo2022Post}. 

In this paper we present a photonic interface designed to connect a single-ion quantum memory with the telecom C-band. Our interface combines a $^{40}$Ca$^+$-resonant, cavity-enhanced SPDC photon-pair source at 854\,nm in interferometric configuration with highly efficient polarization-preserving QFC to 1550\,nm and back-conversion to the atomic wavelength. It thereby facilitates bi-directional sender and receiver operations between atomic and photonic quantum bits \cite{Kurz2016, Brekenfeld2020}, as needed for a quantum network environment involving atom-based memories. We present four different setups and verify entanglement for each case by quantum state tomography: First, we characterize the source operation, then the combined system with QFC to generate polarization entanglement between a $^{40}$Ca$^+$-resonant photon and a telecom photon. In the third step we show entanglement distribution over 20\,km of fiber. Finally we demonstrate the preservation of high-quality entanglement after transmission over 40\,km of fiber and consecutive back-conversion of the telecom photon to the $^{40}$Ca$^+$-wavelength. We thus show for the first time the simultaneous conversion and back-conversion by a single bi-directional frequency converter. Even for the longest distance and bi-directional conversion the entanglement fidelity remains larger than 95\% (98\%) without (with) background correction.

\section{Results}

\subsection{SPDC Photon Pair Source}

\begin{figure}
    \centering
    \includegraphics[width=0.47\textwidth]{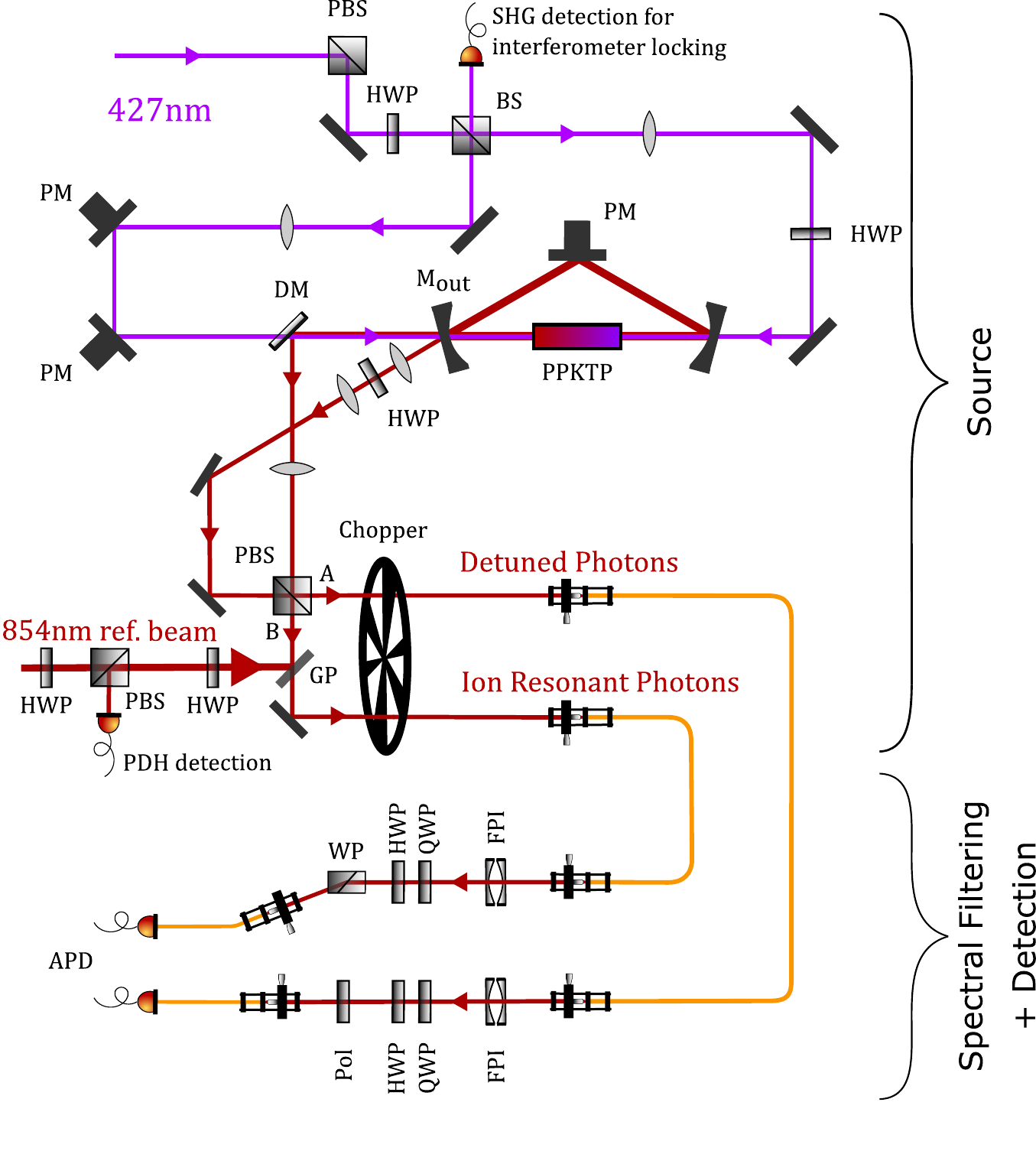}
    \caption{\textbf{Schematic of the photon pair source including filtering and detection.} SPDC source in interferometic configuration with PDH locking beams. Stabilization of the resonator and the interferometer is achieved by piezo driven mirrors (PM). During the stabilization runs, the photons are switched off by an optial chopper in order to protect the APDs. HWP/QWP: half-/quarter-wave plate, PBS: polarizing beamsplitter, BS: non-polarizing beamsplitter, DM: dichroic mirror, Pol: polarizer, WP: Wollaston prism, GP: glass plate, PM: piezo driven mirror, FPI: frequency filter}
    \label{fig:schema_source}
\end{figure}
Our SPDC photon pair source combines the interferometric configuration of \cite{Fiorentino2004, Fedrizzi2007} with cavity enhancement, enabling the generation of photons with very high pair rate, narrow linewidth and high-fidelity, offset-free entanglement simultaneously. Compared to an earlier version of the source \cite{Arenskoetter2017,KuceraPhD}, we also optimised the locking stability and the output mode structure. The experimental setup is shown in Fig.~\ref{fig:schema_source} and will be discussed in the following. 

In order to address the D$_{5/2}$-P$_{3/2}$ transition in $^{40}$Ca$^+$ at 854\,nm, we use a frequency-stable laser at 427 nm to pump the SPDC process in a periodically poled KTP crystal with type-II phase-matching. The polarized pump light is split on a non-polarizing 50:50 beam splitter (BS), and the two beams are coupled into the non-linear crystal in opposite directions. The crystal is placed inside a signal- and idler-resonant 3-mirror ring resonator (FSR$_H=2\pi\cdot1.85\,$GHz, FSR$_V=2\pi\cdot 1.83\,$GHz). Down-converted photons from both pump directions leave the resonator at the same outcoupling mirror, $M_\text{out}$, under different angles. The polarizations of the signal and idler photons in one of the output arms are interchanged by a half-wave plate. Then, the photons of both output arms are overlapped on a polarizing beam splitter (PBS). This arrangement erases all distinguishability between the two photons of a pair. Moreover, it avoids the background of unsplit pairs that is inherent in single-direction SPDC generation. The photonic 2-qubit state at 854\,nm produced by the source is a polarization Bell state \cite{Braunstein1992}
\begin{equation}
    \ket{\Psi} = \frac{1}{\sqrt{2}}\left(\ket{H_\text{A} V_\text{B}}-e^{i\varphi}\ket{V_\text{A} H_\text{B}}\right)
    \label{eq:photonState}
\end{equation}
where $A$ and $B$ refer to the output ports of the PBS.

As shown in Fig. \ref{fig:schema_source} we use a chopped locking beam, injected via a glass plate, to stabilize one mode of the resonator (the H-polarized mode) to the ion transition by the Pound-Drever-Hall (PDH) technique. The frequency of the corresponding V-polarized mode is shifted by $\sim 480$\,MHz because of the birefringence of the crystal. Due to the conversion bandwidth of the crystal (200 GHz) and the small polarisation dispersion of the resonator, several pairs of modes at different frequencies exhibit cavity-enhanced SPDC (see Supplement). Filtering out the ion-resonant mode and its partner mode is effected by two monolithic Fabry-Pérot filters (FPI), one in each output. The filters are described in more detail in the Methods section. 

The SHG light that is produced by the locking beam is detected at the BS that splits the pump beam and is used for the stabilization of the Mach-Zehnder-type interferometer formed between this BS and the PBS that combines the SPDC output arms. By tuning the phase of this interferometer, we tailor the phase $\varphi$ of the photonic 2-qubit state, Eq.~\eqref{eq:photonState}. For the experiments described below, we set this phase to $\varphi=270^{\circ}$. A more detailed account of the relation between $\varphi$ and the interferometer phase is given in the Supplement.

With the described setup we reach a generated photon pair rate of $R_{\text{pair}}= 4.7\cdot10^4\frac{\text{pairs}}{\text{s mW}}\times P_{\text{pump}}$. In the experiments below, we used a maximum pump power of $P_{\text{pump}}=20$\,mW. 
The SPDC photons are available, after spectral filtering and fiber coupling, with 31\% (21.4\%) efficiency in port A (B).

For analysis of the photonic polarization state, we use a projection setup consisting of half-wave plate, quarter-wave plate, and polarizer (film polarizer or Wollaston prism) in each output arm. As single-photon detectors we use two APDs (\textit{Excelitas Technologies}) with approximately $10~\text{s}^{-1}$ dark counts. Detected photons are time-tagged, and time-resolved coincidence functions between the two arms are evaluated for various combinations of polarization settings.

\begin{figure}
    \centering
    \includegraphics[scale=1]{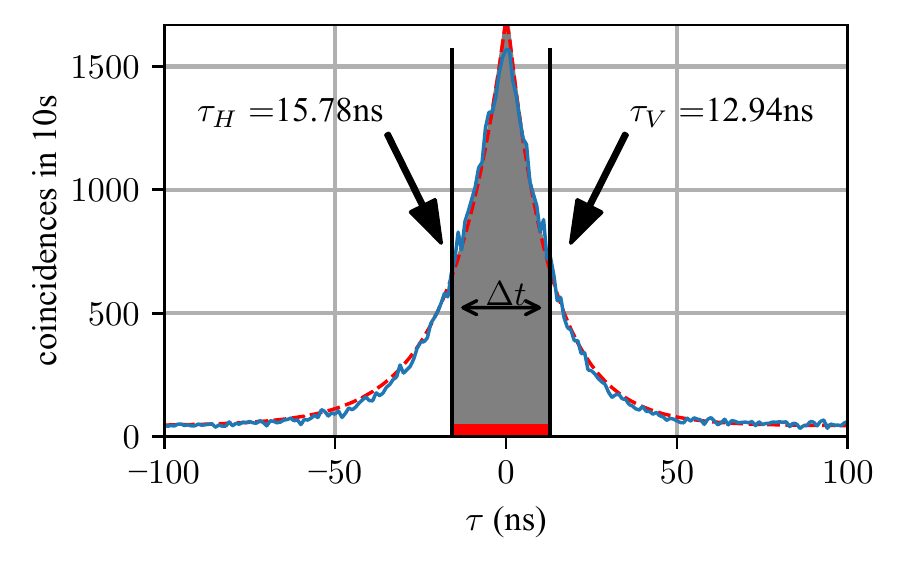}
    \caption{\textbf{Photon wave packet}. Measured signal-idler coincidence function with a bin size of 1 ns. The red dashed line is a double exponential fit to the data.}
    \label{fig:photonSBR}
\end{figure}

For the characterization of the temporal shape of the photon wavepacket, we measure the polarization-indiscriminate coincidence function between the photons of the two output arms \cite{Glauber1963}. Fig.~\ref{fig:photonSBR} shows the result, measured by summing the coincidences for the four settings HV, VH, HH, and VV. The decay times (or wave packet widths) of the two photons differ slightly because of different loss of the two polarization modes in the cavity: we get values of $\tau_H=15.78$\,ns for the H-polarized photons and $\tau_V=12.94$\,ns for the V-polarized photons. 

An important figure of merit is the signal-to-background ratio (SBR) of the coincidence functions. The grey-shaded area in Fig.~\ref{fig:photonSBR}, representing the coincidences in the time-window $\Delta t$ around zero, is taken as the signal ($S$). The background ($B$) is determined using the same time window size but at a delay $\tau>150$\,ns (red shaded area). For our setup the background is dominated by accidental coincidences, originating from the temporal overlap of the generated photons. 

Theoretically, the number of polarization-indiscriminate coincidence counts in a time interval $T$ and for a given pair rate $R_\text{pair}$ is given by 
\begin{equation}
    S = \eta_1 \eta_2 R_\textrm{pair} (1-e^{-\Delta t/2\tau_\text{mean}}) T~,
    \label{eq:Signal}
\end{equation}
where $\eta_1$ and $\eta_2$ are the detection efficiencies for the two outputs, and $\tau_\text{mean}=\frac{1}{2}(\tau_H+\tau_V)$ is the mean wavepacket width of the photons. The theoretical value for the background is  
\begin{equation}
    B = \eta_1 \eta_2 R_\text{pair}^2 \Delta t\,T~.
    \label{eq:Bgd}
\end{equation}
Hence the SBR of the source is expected to be given by \cite{KuceraPhD} 
\begin{equation}
    \text{SBR} = \frac{S}{B} = \frac{1-e^{-\Delta t/2 \tau_\text{mean}}} {R_\text{pair} \Delta t}
    \label{eq:SBR}
\end{equation}

\begin{figure}
    \centering
    \includegraphics[scale=1]{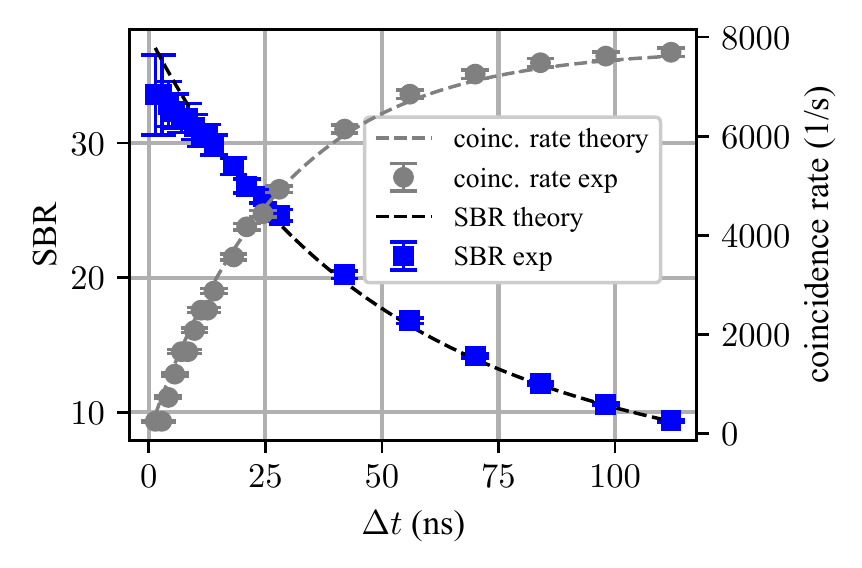}
    \caption{\textbf{Signal-to-background ratio and coincidence rate of the photon pair source}. The SBR and the coincidence rate are plotted in dependence of the coincidence time window $\Delta t$, for a pump power of 20 mW. The error bars are calculated from the Poissonian noise of the measured coincidences. The dashed lines show the theoretical calculations according to Eq.~\eqref{eq:SBR} for the SBR and and Eq.~\eqref{eq:Signal} for the coincidence rate, with $R_\text{pair}$ the only fit parameter.}
    \label{fig:powerSBR}
\end{figure}

Fig.~\ref{fig:powerSBR} shows a comparison between this theoretical expression and our measurement, for fixed pump power of 20\,mW and variable coincidence window $\Delta t$. When $\Delta t$ is similar to the 1/e width of the photon wavepacket, we reach an SBR of $\sim$30, whereas for a coincidence window that covers 99.97\% of the photon, we find an SBR of $\sim$10.

Fig.~\ref{fig:powerSBR} also shows how the coincidence rate varies with $\Delta t$. From fitting Eq.~(\ref{eq:Signal}) to the data, with $\eta_{1,2}$ independently measured, we derive the pair rate $R_\text{pair}$ as mentioned before. 

The SBR enters into the maximally reachable fidelity of the photon pair state to the ideal Bell state of eq. \eqref{eq:photonState}. The fidelity of a measured (i.e., reconstructed) density matrix $\rho$ to a given state $\ket{\Psi}$ is calculated as 
\begin{equation}
    F = \bra{\Psi} \rho \ket{\Psi} = \frac{\frac{1}{4} + F_{\text{w/o BG}} \cdot \text{SBR}} {1+\text{SBR}}
    \label{eq:theoFideltiy}
\end{equation}
where $F_{\text{w/o BG}}$ corresponds to the fidelity of the photonic state when the entire background is subtracted \cite{KuceraPhD}. This formula is used below for the theoretical curves in Fig.~\ref{fig:FidelitySBR}.

\vspace*{.5cm}

\subsection{Polarization Preserving Frequency Conversion}

\begin{figure*}
\begin{subfigure}[c]{0.45\textwidth}
\caption{}
    \includegraphics[width=\textwidth]{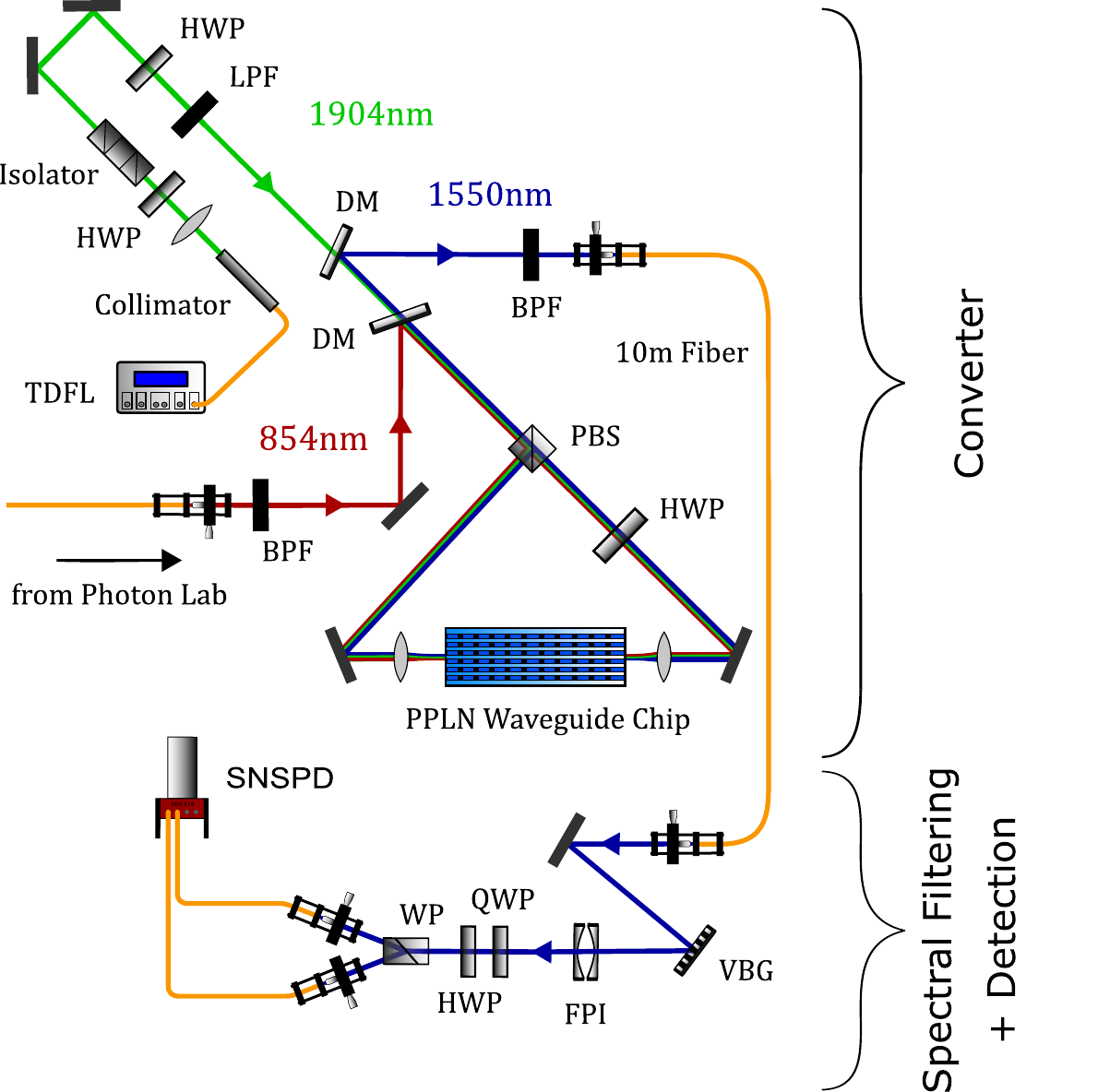}
    \label{fig:Schema_Konverter}
\end{subfigure}
\begin{subfigure}[c]{0.45\textwidth}
\caption{}
    \includegraphics[width=\textwidth]{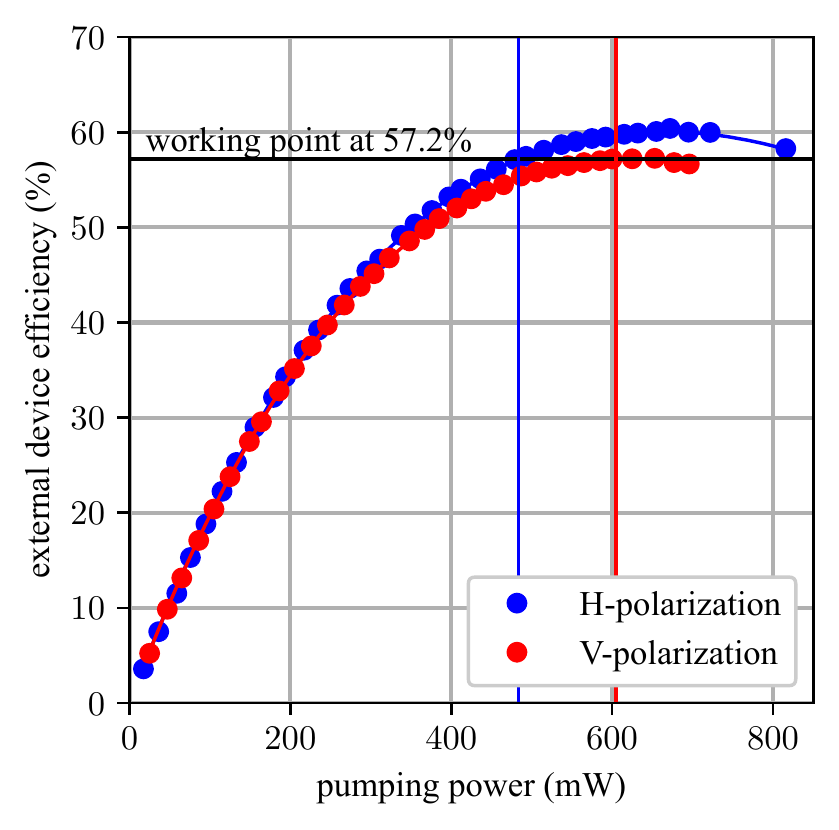}
    \label{fig:Konversionseffizienz}
   \end{subfigure}
   \caption{\textbf{a) Schematic quantum frequency conversion setup.} Details are found in the main text. TDFL: Thulium-doped fiber laser, LPF: low-pass filter, BPF: band-pass filter, VBG: volume-Bragg grating, DM: dichroic mirror, HWP: half-wave plate, SNSPD: superconducting nanowire single photon detector; \textbf{b) Device efficiency of the converter for different pump powers and input polarizations.} The maximum efficiencies of the H- and V-polarizations are slightly different.}
\end{figure*}

The setup for the polarization-preserving frequency conversion (Fig.~\ref{fig:Schema_Konverter}) is located in a second lab and is connected to the photon-pair source via 90\,m of fiber. We use difference frequency generation (DFG) in a nonlinear PPLN waveguide to convert the 854\,nm photons to the telecom C-band at 1550\,nm \cite{Krutyanskiy2017}. As the DFG efficiency in the waveguide is strongly polarization dependent, the setup is realized in a Sagnac configuration \cite{Ikuta2018, vanLeent2020}. First, the 854\,nm signal is filtered by an input bandpass filter to prevent background photons being coupled back into the source setup and overlapped with the strong, diagonally polarized pump field at 1904\,nm on a dichroic mirror. Both fields are then split into their H- and V-polarization component on a PBS. To achieve polarization preserving operation, the H-components are rotated to V by an achromatic waveplate inside the Sagnac loop (HWP). All beams are now V-polarized for optimum conversion and are coupled in a counterpropagating way into the waveguide. After exiting the waveguide, the converted 1550-nm light from the original V-component also passes the achromatic waveplate and is thereby rotated to H. Finally, the two converted polarization components are coherently overlapped on the PBS, which closes the Sagnac loop. The light then passes the dichroic mirror a second time where unconverted (or back-converted) light at 854\,nm is split off. The same happens to SHG light from the pump that is parasitically generated in the waveguide; it is then blocked by the input bandpass filter. The converted light that is transmitted through the first DM together with the pump light is separated with a second DM and coupled into a telecom fiber, directing it to the detection setup.

Apart from enabling polarization-independent conversion, an additional advantage of the Sagnac configuration is that all fields have the same optical path, such that no phase difference between the split polarization components occurs, and hence no active phase stabilization is needed. At the same time, the configuration facilitates bi-directional conversion without compromising the conversion efficiency. An experimental demonstration of bi-directional operation will be presented below.  

The conversion-induced background in this process mainly originates from anti-Stokes Raman scattering of the pump field \cite{Zaske2011, Krutyanskiy2017, Kuo2018}. As this is spectrally very broad compared to the converted signal, we can significantly reduce the background count rate with a narrowband filtering stage. By combination of a bandpass filter (transmission bandwidth $\Delta \nu_{\text{BPF}} = 1500$\,GHz), a volume Bragg grating (VBG, $\Delta\nu_{\text{VBG}}=25$\,GHz) and a Fabry-Pérot etalon (FSR\,=\,12.5\,GHz, $\Delta\nu_{\text{FPE}}\,=\,250$\,MHz) we achieve broadband suppression outside a 250\,MHz transmission bandwidth. To ensure high transmission through VBG and Etalon, a clean gaussian spatial mode is needed. That is why these filters are included in the detection setup, which is separated from the conversion setup by 10\,m of fiber (or later the fiber link) acting as spatial filter. With this filtering stage, the total conversion-induced background count rate is 24\,s$^{-1}$.

The device efficiency including all losses in optical components and spectral filtering for H and V input polarizations and different pump powers in the corresponding converter arm is shown in Fig. \ref{fig:Konversionseffizienz}. The measurement agrees well with the theoretical curve given by $\eta_\text{ext}(P)= \eta_\text{ext,max}\sin^2(\sqrt{\eta_\text{nor}P}L)$ with the normalized power efficiency $\eta_\text{nor}$ and waveguide length $L$ \cite{Roussev2002}. We achieve a maximum external conversion efficiency of $\eta_\text{ext,max}=\,60.1\%\,(57.2\%)$ for H(V)-polarized input at 660\,mW\,(630\,mW) pump power. The difference is explained by 
slightly different mode overlaps between signal and pump field in the two arms. Since for polarization-preserving operation both efficiencies need to be equal, the pump power in the H-arm is reduced via the two HWPs in the pump laser arm to 485\,mW to match 57.2\% conversion efficiency. To verify equal conversion efficiency for arbitrary polarized input, the process matrix \cite{Chuang2008} was measured with attenuated laser light, resulting in a value for the process fidelity of 99.947(2)\%. Thus the preservation of the input polarization state is confirmed with very high fidelity. Note that in fact the converter rotates the input state by a constant factor $\pi/2$ due to the achromatic waveplate. This could be compensated by an additional waveplate at the output, however the polarization also gets arbitrarily rotated in the input and output fibers, so we compensate for all rotations together by rotating the projection bases in all measurements as described in the Methods section. \\
Further details on the conversion setup as well as the background and process fidelity measurements are found in the supplement.

\subsection{Experimental results}

\begin{figure}
    \centering
    \includegraphics[width=0.45\textwidth]{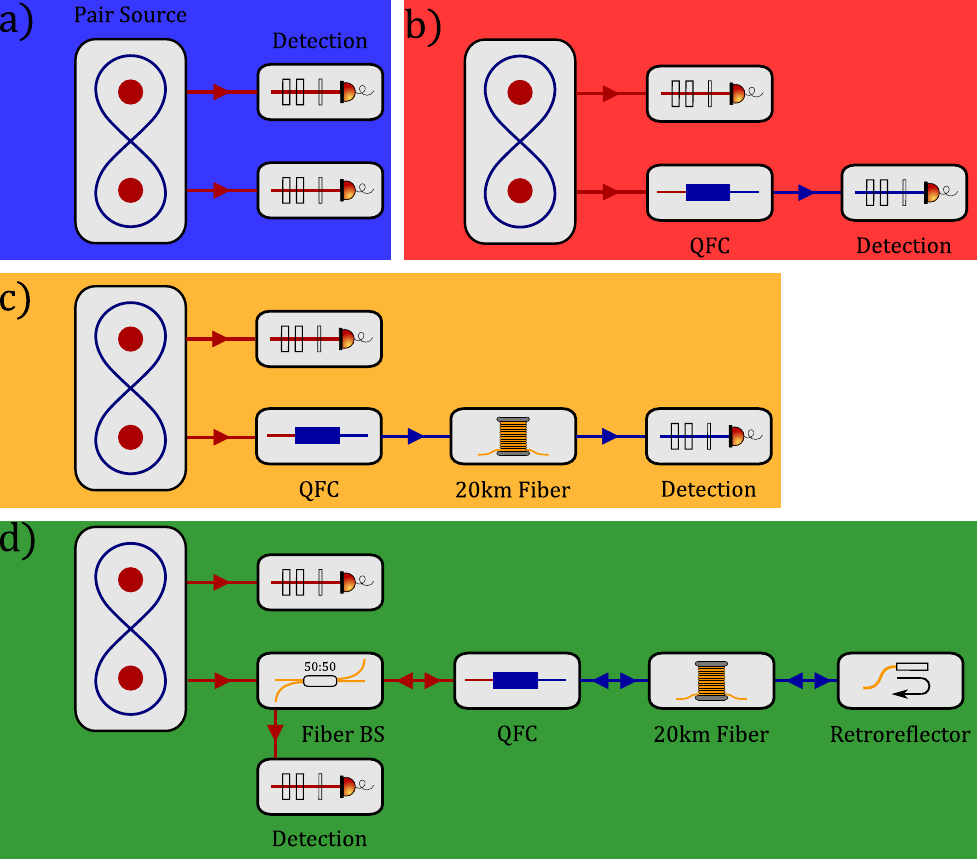}
    \caption{\textbf{Overview of the different experimental configurations.} \textbf{a)} The output state is measured directly at the source. \textbf{b)} One of the photons is frequency-converted before its detection. \textbf{c)} 20\,km of fiber is inserted between conversion and detection. \textbf{d)} The 20\,km fiber gets terminated with a retro-reflector. The back reflected photons get back-converted after 40 km of fiber and are extracted with a 50:50 fiber beamsplitter before their detection. The background colors correspond to the colors of the data points in Fig. \ref{fig:FidelitySBR}.}
    \label{fig:setup_complete}
\end{figure}

To assess the performance of our quantum photonic interface we analyze the photon pair states via quantum state tomography \cite{Kwiat2001} for four different configurations, as sketched in Fig.~\ref{fig:setup_complete}a-d. 

In the first configuration (Fig.~\ref{fig:setup_complete}a) and for calibration purposes, we measure the output state of the photon pair source itself, i.e., without conversion. In the next configuration (Fig.~\ref{fig:setup_complete}b), we include the frequency converter in the detuned output arm of the photon pair source (arm A in Fig.~\ref{fig:schema_source}). Tomography is then performed on the pairs of converted and unconverted photons. 

In the third configuration (Fig.~\ref{fig:setup_complete}c), we extend the fiber link between converter and detection setup to 20\,km in order to evaluate the influence of fiber transmission. 
We emphasize that the transmission of photons at a wavelength of 854 nm via a suitable single mode fiber would suffer about -70 dB fiber attenuation \cite{Nufern2020} whereas conversion to 1550 nm and using a low-loss telecom fiber results in a total fiber attenuation of -3.4\,dB \cite{Corning2020}.

Finally, in the fourth configuration (Fig.~\ref{fig:setup_complete}d) bi-directional conversion is implemented by terminating the 20\,km fiber with a retroreflector. Thus, the telecom photons are converted back to 854 nm at the second passage of the converter, after 40 km of fiber transmission. To separate the returning back-converted photons from the outgoing ones in the source lab, we use a 50:50 fiber beamsplitter that reflects 50\% of the back-converted photons to the second tomography setup. Note that in this configuration the converter filter stage is not used, but the FPI filter in the 854\,nm tomography setup, as explained in the supplement.\\

\begin{figure*}
    \begin{subfigure}{0.49\textwidth}
        \includegraphics[scale=1]{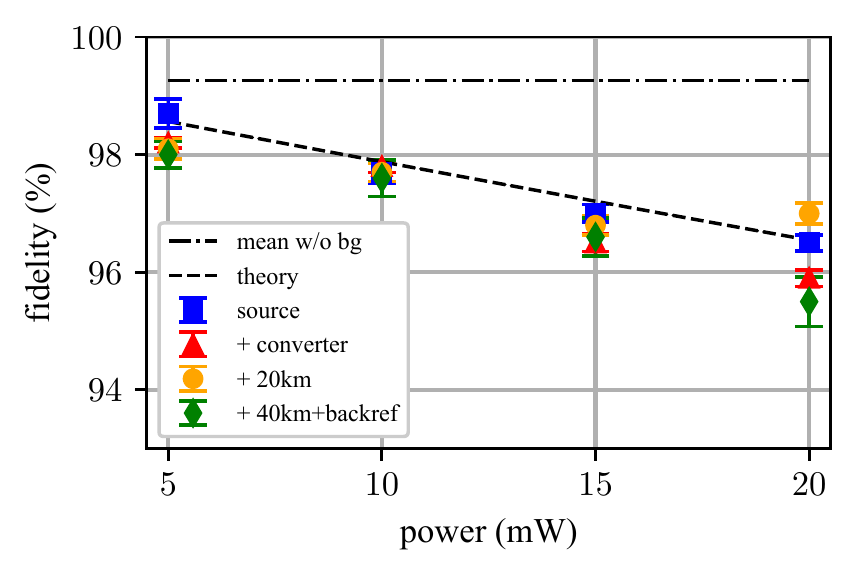}
        \subcaption{Fidelity for a detection window $\Delta t= 1.5\tau_\text{mean}$}
    \end{subfigure}
    \begin{subfigure}{0.49\textwidth}
        \includegraphics[scale=1]{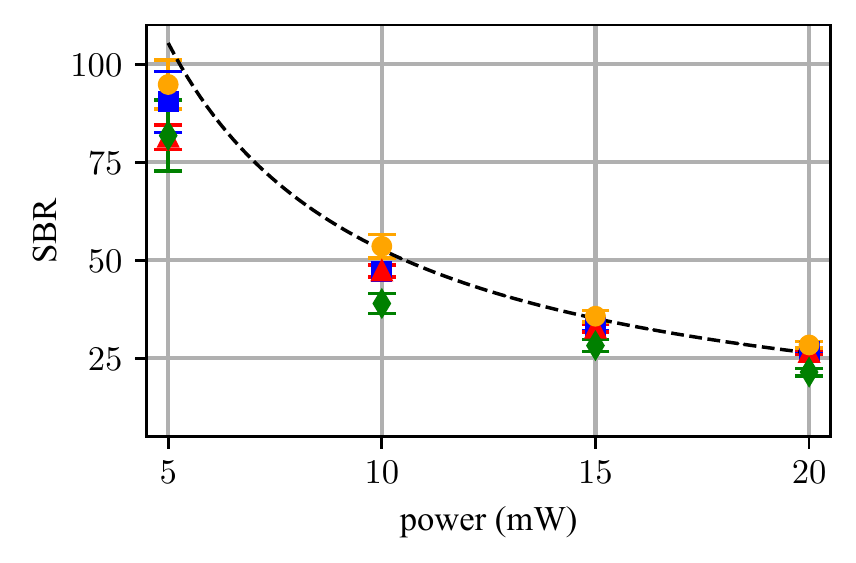}
        \subcaption{SBR for a detection window $\Delta t= 1.5\tau_\text{mean}$}
    \end{subfigure}
    \begin{subfigure}{0.49\textwidth}
        \includegraphics[scale=1]{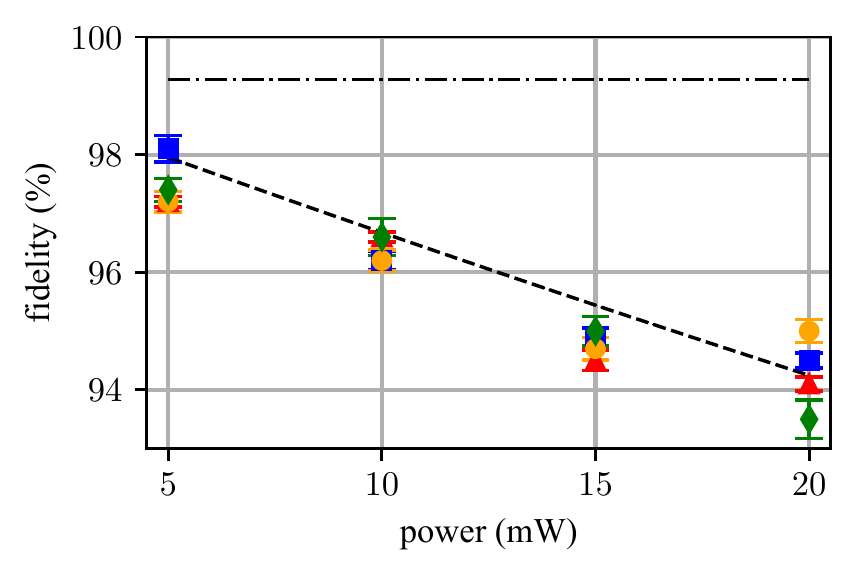}
        \subcaption{Fidelity for a detection window  $\Delta t= 5\tau_\text{mean}$}
        \label{fig:Fidelity1.5tau}
    \end{subfigure}
    \begin{subfigure}{0.49\textwidth}
        \includegraphics[scale=1]{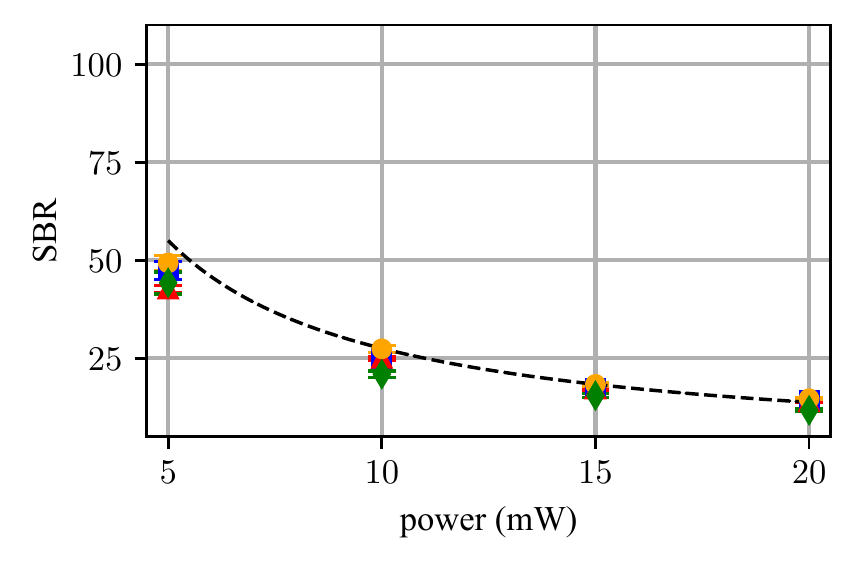}
        \subcaption{SBR for a detection window $\Delta t= 5\tau_\text{mean}$}
        \label{fig:SBR1.5tau}
    \end{subfigure}
     \caption{\textbf{Fidelity and SBR vs.\ SPDC pump power for the four configurations of Fig.~\ref{fig:setup_complete}.} a) and b) show the measurement results for a detection time window of $1.5\tau_\text{mean}$ which corresponds to 78\% of the photon temporal shape. In a) the MLE results are represented by the data points, the dot-dashed line shows the mean background corrected fidelity, while the dashed line shows the theoretical expectation according to eq. \eqref{eq:theoFideltiy}. The uncertainties are derived from a simulation; see Method section for further details.
     b) shows the comparison of the SBR for the four different measurement settings. The dashed line is calculated for the given pair rate according to eq. \eqref{eq:SBR}. The error bars are calculated by including the $\sqrt{N}$-noise of the measured coincidences \cite{Altepeter2005}.
     c) and d) show the results when choosing a larger detection window of $5\tau_\text{mean}$ for the evaluation. 
     }
     \label{fig:FidelitySBR}
\end{figure*}

In all four configurations, we use source pump powers of 5\,mW, 10\,mW, 15\,mW and 20\,mW. The detected coincidence rates at 20 mW pump power are: 4168 s$^{-1}$ for setup a), 974 s$^{-1}$ for setup b), 428 s$^{-1}$ for setup c) and 40 s$^{-1}$ for setup d) (also see Supplement).From the time-resolved coincidences in the different detection bases, the density matrix is reconstructed by applying an iterative maximum likelihood estimation (MLE) algorithm \cite{Hradil1997}. From the density matrix, we then calculate all measures that characterize the quantum state, such as fidelity and purity. For each configuration we show a representative density matrix in the supplement. Before performing the tomography, the polarization rotations in both arms, including the converter and the fiber link, were calibrated; details on the procedure are found in the methods part.

The results of the four configurations are summarized in Fig.~\ref{fig:FidelitySBR}.
For all configurations and source pump powers, the fidelities with respect to the maximally entangled state (\ref{eq:photonState}) are above 95\% with and without background correction, for a detection window of $\Delta t = 1.5\,\tau_\text{mean}$ (subfigure a). Thus the entanglement is well-preserved for every scenario of Fig.~\ref{fig:setup_complete}. 
For fixed source pump power, the fidelities of the individual configurations differ only slightly outside the error bars, which confirms the high process fidelity of the converter and the low detrimental effect of the fiber link.
With increasing source pump power, the fidelity decreases, which is expected according to eq.~(\ref{eq:theoFideltiy}). The black dashed line shows the theoretically achievable fidelity for the given SBR, considering only accidental coincidences. This line is expected to be an upper bound to the measured values, as additional effects lead to further reductions: (i) errors in the calibration of the polarisation rotation compensation; (ii) drifts in the polarization rotation of the setup away from the calibration point during the measurements; (iii) fluctuations of the photonic state phase resulting from power fluctuations of the locking signal. 

The SBR for each individual measurement is shown in Fig.~\ref{fig:FidelitySBR}b. Here the error bars are calculated by taking Poissonian noise into account. As expected from eq.~(\ref{eq:SBR}), the SBR decreases for increasing pump power of the pair source. Again we only see minor differences between the four configurations, as the conversion-induced background rate is negligible compared to the pair rate and source background rate in the relevant time window. For the back-conversion part, the SBR is consistently lower, which is explained by the use of the FPI filter instead of the converter filtering stage.


When choosing a larger detection window, fidelities and SBR decrease due to the temporal overlap of the photons. The results for a detection window of $\Delta t=5\,\tau_\text{mean}$, corresponding to 91\% of the photon, are shown in Fig.~\ref{fig:FidelitySBR}c-d. Due to the lower SBR, here the fidelities are lower by a few percent, while the detected pair rate is increased by a factor of 1.7.

\section{Discussion}
In summary, we have presented and characterized a combined system of entangled photon pair source and bi-directional quantum frequency converter and use it for entanglement distribution over a fiber link. Pair rate and entanglement fidelity of the source are near the theoretical optimum, while efficiency, polarization fidelity and noise performance of the converter are among the best reported values \cite{Krutyanskiy2017,Ikuta2018,Kaiser2019}. The system is tailored as interface to the $^{40}$Ca$^+$ single-ion quantum memory. 

The presented measurements demonstrate the preservation of photon entanglement with high fidelity through up to 2 conversion steps and distribution over up to 40\,km of fiber. The high-efficiency bi-directional quantum frequency conversion adds low background to the photon pair source and thus enables a proof-of-principle demonstration of a quantum photonic interface suitable to connect remote quantum network nodes based on $^{40}$Ca$^+$ single-ion memories. 

Several future improvements are readily identified. The coincidence rates can be enhanced by replacing lossy fiber-fiber couplings by direct splices and by replacing the fiber beam splitter by an optical circulator. Given the low unconditional background rate of the QFC process, we can further increase photon count rates by lowering the finesse of the monolithic Fabry-Pérot filters, which results in higher transmission. The conversion fidelity can be improved near the theoretical maximum by repeating the polarization calibration more frequently in a fully automated procedure. Furthermore, a redesign of the source-cavity itself would make it possible to shorten the photons to 8\,ns, which corresponds to the transition linewidth of 23\,MHz between the D$_{5/2}$ and P$_{3/2}$ state. Due to less overlap between two subsequent photons, the SBR and the fidelity would improve as well.

Recent developments in quantum networking operations with single ions \cite{Krutyanskiy2022entanglement,Krutyanskiy2022telecom,Drmota2022}, together with progress in ion-trap quantum processors \cite{Pogorelov2021,Noel2022} confirm the importance and the potential of quantum photonic interfaces for quantum information technologies. The interface we demonstrate here enables extension towards further fundamental operations in trapped-ion-based quantum networks. In particular, bi-directional QFC allows for teleportation between remote quantum memories, or for their entanglement via direct exchange of a photon (see e.g. \cite{Luo2022}), employing heralded absorption \cite{KuceraPhD, Kurz2016} of a back-converted telecom photon. 

\section{Methods}

\subsection*{State characterization}

We perform full state tomography to reconstruct the photonic two-qubit state. We project each photon to one of the six polarization basis states (horizontal, vertical, diagonal, anti diagonal, left circular, or right circular), which results in 36 possible measurement combinations. In order to reconstruct the 2-qubit density matrix, 16 combinations are sufficient. We therefore measure two-photon correlations in 16 independent polarization settings \cite{James2001} and reconstruct the density matrix by applying a maximum likelihood algorithm to the count rates inside the detection window \cite{Hradil1997}. We infer from the density matrix all characteristic measures such as purity and fidelity. 
For calculating the error bars, we use a Monte Carlo simulation \cite{Altepeter2005} where we run the algorithm repeatedly on randomly generated, Poisson-distributed coincidence rates based on the measured rates. From the distribution of the resulting fidelities, we estimate the error bars.

\subsection*{Polarization bases calibration}

Polarization rotation due to the birefringence of optical components in the setup, in particular the fibers, has to be accounted for to ensure projection onto the correct basis states. We model the rotation in each arm by a unitary rotation matrix $M$ that transforms the input polarization $\lambda_{\text{in}}$ to an output polarisation $\lambda_{\text{out}} = M\lambda_{\text{in}}$. Non-unitary effects such as polarization-dependent loss play only a negligible role on the final state fidelity.

 By blocking one pump direction of the source, we deterministically generate photons with linear (H) polarization. Additionally, we can insert a quarter-wave plate directly behind the source to rotate the polarization to R. We then perform single-qubit tomography with the detection setups to measure the rotated polarization states. From this we calculate the rotation matrix $M$, and the projection bases are rotated accordingly.

Depending on the experimental configuration (Fig.~\ref{fig:setup_complete}), we measure the rotation matrix less or more frequently during the experiments: for the first two configurations, short fibers are used, thus the polarization is very stable over time and is measured only at the beginning of the experiment. In the last two configurations, although the fiber spool is actively temperature stabilized to minimize polarization drifts, the rotation matrix is measured every 60 minutes. 

\subsection*{Fabry-Pérot filter}
The Fabry-Pérot filters which we use for filtering the 854\,nm photons to a single frequency mode are built from single 2 mm thick NBK-7 lenses. Both sides have a high-reflectivity coating ($R=0.9935$). This results in a finesse of 481, a FSR of $\sim 50$~GHz and a FWHM of $\Delta\nu\approx104$ MHz, which is sufficient for filtering a single mode of the SPDC cavity (FSR $\sim 1.84$~GHz). Frequency tuning over a whole FSR is possible by changing the temperature in the range between 20°C and 70°C. The stabilization of the temperature with a precision of 1 mK (corresponding to 3 MHz) is sufficient for stable operation; no further active stabilization is needed.

\section{Data availability}
The underlying data for this manuscript is openly available in Zenodo at \url{https://doi.org/10.5281/zenodo.7313581}.
The evaluation algorithms are available from the corresponding author upon reasonable request.

\section{References}
\bibliography{paper}

\section{Acknowledgements}
We acknowledge support by the German Federal Ministry of Education and Research (BMBF) through projects Q.Link.X (16KIS0864) and QR.X (16KISQ001K).

\section{Author contributions}
E.A., T.B., S.K., and M.B. designed the experiments. E.A. and T.B. performed the
experiments. E.A. and T.B. analysed the data. E.A., T.B., S.K. and M.B.
prepared the experimental setup. E.A. and T.B. wrote the paper with
input from all authors. J.E. and C.B. supervised the project.

\section{Competing interests}
The authors declare no competing interests.


\end{document}


\title{Supplement: Telecom Quantum Photonic Interface for a $^{40}$Ca$^+$ Single-Ion Quantum Memory}

\author{Elena Arensk\"otter}
\author{Tobias Bauer}
\author{Stephan Kucera}
\affiliation{Experimentalphysik, Universit\"at des Saarlandes, 66123 Saarbr\"ucken, Germany}
\author{Matthias Bock}
\thanks{} 
\affiliation{Experimentalphysik, Universit\"at des Saarlandes, 66123 Saarbr\"ucken, Germany}
\affiliation{present address: Institut für Quantenoptik und Quanteninformation,
\"Osterreichische Akademie der Wissenschaften, Technikerstraße 21a, 6020 Innsbruck, Austria}
\author{J\"urgen Eschner}
\email{juergen.eschner@physik.uni-saarland.de}
\author{Christoph Becher}
\email{christoph.becher@physik.uni-saarland.de}
\affiliation{Experimentalphysik, Universit\"at des Saarlandes, 66123 Saarbr\"ucken, Germany}

\date{\today}

\maketitle

\setlength{\parindent}{0pt}
\renewcommand{\figurename}{Supplementary Figure}
\renewcommand{\tablename}{Supplementary Table}
\setcounter{figure}{0}
\subsection*{Supplementary Methods 1: Source}

\subsubsection*{Further details to the experimental setup}
In the following we give further details on the setup of the source. The schematic of the source is shown in Figure 1. The 427\,nm pump light is polarized by a polarizing beam splitter (PBS) and is rotated to the correct polarization by a $\lambda/2$ wave plate (HWP). This ensures that most of the light is used to generate photon pairs, because only one polarization is converted. After that, the pump light is split on a non-polarizing beam splitter (BS). The reflected and the transmitted beams pump the resonator from opposite sides. 
Photons generated from the transmitted pump beam leave the resonator at mirror $M_{out}$ under 0° and are separated from the pump by a dichroic mirror (DM). After that, we send the photons to a PBS. 
The photons generated from the other pump beam leave the resonator under 30°, passing two lenses for beam-shaping and an additional HWP, which interchange the polarization of signal and idler photon.
The output state can then be written as
\begin{equation}
    \ket{\Psi_{\text{out}}} = \frac{1}{\sqrt{C_1^2+C_2^2}}(C_1 \ket{H_AV_B}-C_2e^{i\varphi}\ket{V_AH_B})
    \label{eq:state}
\end{equation}
with weight factors $C_1$ and $C_2$.
To generate a Bell state,  both pump directions need to be balanced in means of coincidence counts, resulting in C1=C2. We achieved the balancing of coincidence counts with an additional HWP in one of the pump beams by slightly misaligning the polarization of one of the beams.

\subsubsection*{Photon spectrum}
\begin{figure}[h]
    \centering
    \includegraphics[width=0.45\textwidth]{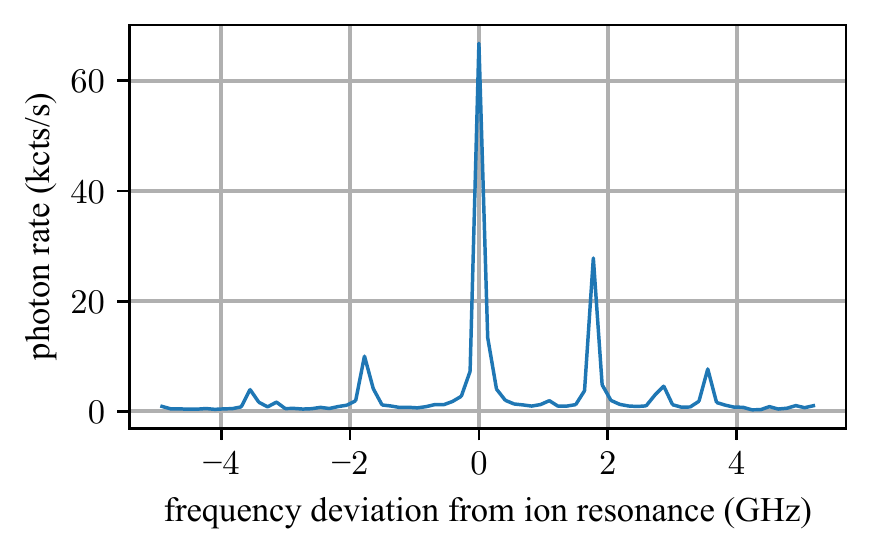}
    \caption{\textbf{Measured cluster structure of the photon pair source} The plots shows the frequency resolved photonic cluster of the photon pair source. The counts are recorded while scanning the FPI filter over the resonance.}
    \label{fig:cluster}
\end{figure}

The small polarisation dispersion of the cavity together with the conversion bandwidth of the crystal ($\sim200\,$ GHz) leads to a cluster structure in the produced photon spectrum. This structure was measured by scanning the FPI over a range of $\pm5\,$GHz around the resonance, while measuring the count rate behind it. Supplementary Fig. \ref{fig:cluster} shows the result. We get five peaks in the spectrum with a spacing of $\sim1.8\,$GHz, which is the FSR of the cavity. We infer from the area under the peaks that we get around 60 \% of the power in the central peak.

\subsubsection*{Stabilization}
Both, the resonator and the interferometer are actively stabilized. For the resonator, we use the PDH technique. Via a glass plate (GP) we inject ion-resonant 854\,nm light from a reference laser through the outcoupling mirror M$_{\text{out}}$. By choosing diagonal polarization for the light, it couples to both directions of the resonator, which is necessary for generation SHG light in both directions. The HWP again ensures that both directions have the same polarization. The reflection from the resonator is then detected on a photo-diode. Feedback is applied via a piezo-movable mirror.

The interferometer is stabilized in the following manner: the light used for the PDH-lock also generates a small amount of SHG light in both pump directions such that a Mach-Zehnder interferometer is formed between the red PBS and the blue BS. An APD in current mode is attached to the free port of the BS to detect the fringes. Feedback is applied to a pair of piezo-movable mirrors in one of the pump arms. The measured relation between the interferometer phase and the reconstructed photonic state phase $\varphi$ (eq. \eqref{eq:state}) is shown in Supplementary Figure \ref{fig:statePhase}. 
At the turning points of the fringe, i.e. the gray shaded areas, no stabilization is possible. This is the reason why we use a phase of 270$^\circ$ instead of $0^\circ$ or $180^\circ$ which would result in the conventional $\Psi^+$ or $\Psi^-$ Bell state. The offset between the state phase and the interferometer phase depends mainly on the path length difference of the two interferometer paths. For the case of equal paths the resulting phase is around 270$^\circ$.
 
To protect the APDs from laser light, both locks are running in a sequential chopped mode. All beams (pump light, lock light 854\,nm, SHG light, and photons) pass the same chopper to synchronize the operations.

\begin{figure}
    \centering
    \includegraphics[width=0.45\textwidth]{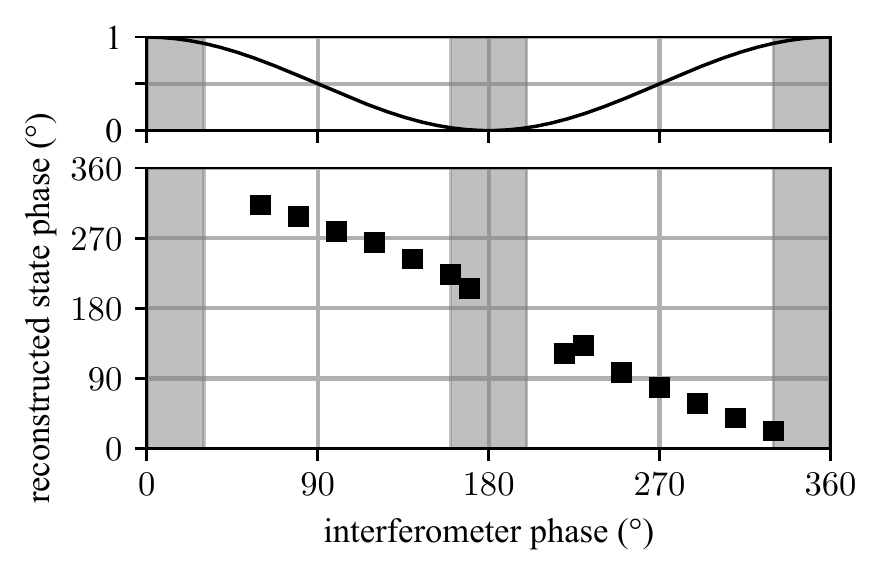}
    \caption{\textbf{Relation between interferometer phase and state phase.} The plot shows the measured state phase for different interferometer phases. In the upper plot the theoretical interferometer fringe is displayed. The grey shaded areas indicate regions where no locking is possible.}
    \label{fig:statePhase}
\end{figure}

\subsection*{Supplementary Methods 2: Converter
}

\subsubsection*{Waveguide Coupling}

For optimal conversion efficiency, signal and pump field are focused onto their respective fundamental WG mode. For this, we employ aspheric lenses (AL), which at the same time couple out the converted light. Due to chromatic aberration, the focal length is different for all three wavelengths. Thus it is not possible to couple all three fields simultaneously when using collimated beams. To deal with this, we generate non-collimated beams with appropriate beam parameters at the WG coupling lens by varying the distances between the WG coupling lens and the fiber coupling lenses as shown in Supplementary Figure \ref{fig:WGCoupling}. As the pump laser has an output collimator, we use an additional spherical lens (SL) for beam shaping.
\begin{figure}
    \centering
    \includegraphics[width=0.45\textwidth]{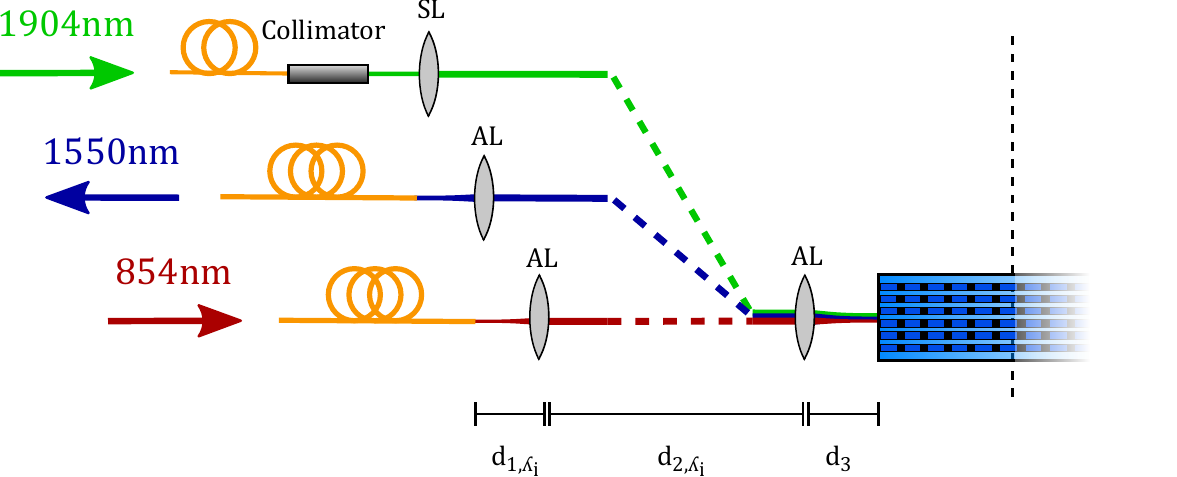}
    \caption{\textbf{Schematic of the distances to be optimized for high coupling efficiency. }}
    \label{fig:WGCoupling}
\end{figure}
Instead of finding the optimal distances empirically we optimize them numerically. For that we first calculate the fundamental WG mode for all 3 wavelengths. From the initial beam parameters given by the fiber MFDs and the beam profile of the pump laser we calculate for each wavelength the field distribution at the WG facette as a function of the three distances $d_{1,\lambda_i}, d_{2,\lambda_i}, d_{3}$ by matrix optics for each wavelength. The overlap between the field distribution and the WG fundamental mode then gives a measure for the coupling efficiency. We then numerically maximize the overlap for all 3 wavelengths by varying $d_{1,\lambda_i}, d_{2,\lambda_i}, d_{3}$ within bounds reasonable for the setup. Note that $d_3$ is a shared parameter as all fields are coupled to the WG with the same lens. Due to the large diameter pump beam, it was not possible to achieve maximum coupling efficiency for all 3 fields simultaneously. In the optimization we therefore weight the pump field coupling efficiency less than the single photon fields as we have enough pump power available. The simulation was repeated for a number of commercially available lenses where the chromatic focal shift data was available. The best combination of distances and focal lengths results in a coupling efficiency of 99.5\% for 854\,nm and 1550\,nm respectively and 91.9\% for 1904\,nm. The achieved experimental values are: For 854\,nm 96.2\%(97.1\%) for the H(V)-arm, for 1904\,nm  89.4\% (88.9\%), for the 1550\,nm fiber coupling we get 93.5\% where reflection losses at the fiber facets are included. Other possible factors leading to deviations from the simulated results are aberrations in the coupling lenses and non-perfect WG geometries which distort the WG modes. For further details see \cite{BockPhDS}.

\subsubsection*{Frequency Converter Losses}

\begin{table}
\caption{ Transmission and Efficiencies of individual components of the conversion setup}
\label{tab:Efficiencies}
\begin{tabular}{c c c}	& H & V \\ 
	\hline
	\hline
	Optical elements transmission, 854 & 92.3\% & 93\% \\ 
	AL transmission, 854 & \multicolumn{2}{c}{93\%} \\ 
	WG coupling efficiency & 96.2\% & 97.1\% \\ 
	AL transmission 1550 (3x) & \multicolumn{2}{c}{94.4\%} \\ 
	Bandpass filter transmission & \multicolumn{2}{c}{96\%} \\ 
	Fiber in-/out coupling efficiency & \multicolumn{2}{c}{93.5\%} \\ 
	VBG transmission & \multicolumn{2}{c}{98.5\%} \\ 
	Etalon transmission & \multicolumn{2}{c}{93.4\%} \\
	Internal efficiency + Opt. el. tr., 1550\qquad\qquad & 93.3\% & 87.4\% \\
	\hline
	\vspace{1cm}
	Device Efficiency & 60.1\%& 57.2\%\\
	\end{tabular}
	
\end{table} 
The total device efficiency of the converter is composed of the internal efficiency of the conversion process, coupling efficiencies, and losses induced by all optical elements. A detailed list of transmissions and efficiencies of all components is shown in Supplementary Table I. The 854\,nm signal passes the aspheric fiber coupling lens, a dielectric mirror, a dichroic mirror, the PBS and two silver mirrors. In total, the setup transmission up to the WG coupling lens is 92.3\%(93\%) for the H(V)-arm. Although the WG coupling lens is custom AR-coated for all three wavelengths, it shows a transmission of only 93\% for 854\,nm. It was not possible to also measure the total setup transmission for the converted 1550\,nm light. As the converted light is overlapped with the strong pump laser behind the WG either a dichroic mirror to reflect the pump light out of the 1550\,nm path or a seperate 1550\,nm laser would be needed which weren't available. Only the aspheric lens transmission could be measured for 1550\,nm. The total transmission through all 3 lenses in the 1550\,nm path (WG coupling lens and fiber in and out coupling lenses) is 94.4\%. The bandpass filter shows a transmission of 96\%. The fiber between conversion and detection setup is AR-coated only on the input side, the combined in and out coupling efficieny is 93.5\%. For the VBG and Etalon we get transmissions of 98.5\% and 93.4\% respectively. From these numbers and the device efficiency we can now infer the combined internal efficiency and 1550\,nm setup transmission which is 93.3\%(87.4\%) for the H(V)-arm.
It is apparent, that the device efficiency is not limited by a single element with low transmission, but rather the sum of many components with already low individual loss. For a significant improvement of the device efficiency we therefore need to replace several components.
For future experiments we will therefore use custom dielectric mirrors with reflectivities > 99.5\% for the single photon wavelengths instead of the currently used silver mirrors for WG coupling. The WG coupling lens will be replaced by one with transmission > 99\% for 854\,nm and 1550\,nm. For connecting converter and detection setup, we will use a fiber with AR coating on both ends. Thus, a device efficiency of up to 70\% is within reach.

\subsubsection*{Laser Process Tomography}
To characterize the converter polarization preservation for arbitrary polarized light, quantum process tomography with attenuated laser light is used.
For that, first the polarization rotation matrix of the conversion and detection setup is measured and compensated as described in the Methods section. Here we prepare 37 polarization states at the converter input with a linear polarizer and two waveplates and measure the corresponding output polarization with the single photon detectors. We then prepare the polarizations H,V,D,A,R and L at the converter input and perform quantum state tomography on the output states. From the resulting photon count rates in the different detection basis settings, the process matrix in the Pauli-basis ($\sigma_0$,$\sigma_1$,$\sigma_2$, $\sigma_3$) is reconstructed with a maximum likelihood algorithm \cite{BockPhDS}. The result is shown in Supplementary Figure \ref{fig:Prozessmatrix}. We only get a significant contribution from the first matrix entry, which corresponds to the process fidelity of 99.947(2)\% where the error is calculated with a Monte Carlo simulation assuming Poissonian photon statistics. 

\begin{figure}
    \centering
    \includegraphics[width=0.45\textwidth]{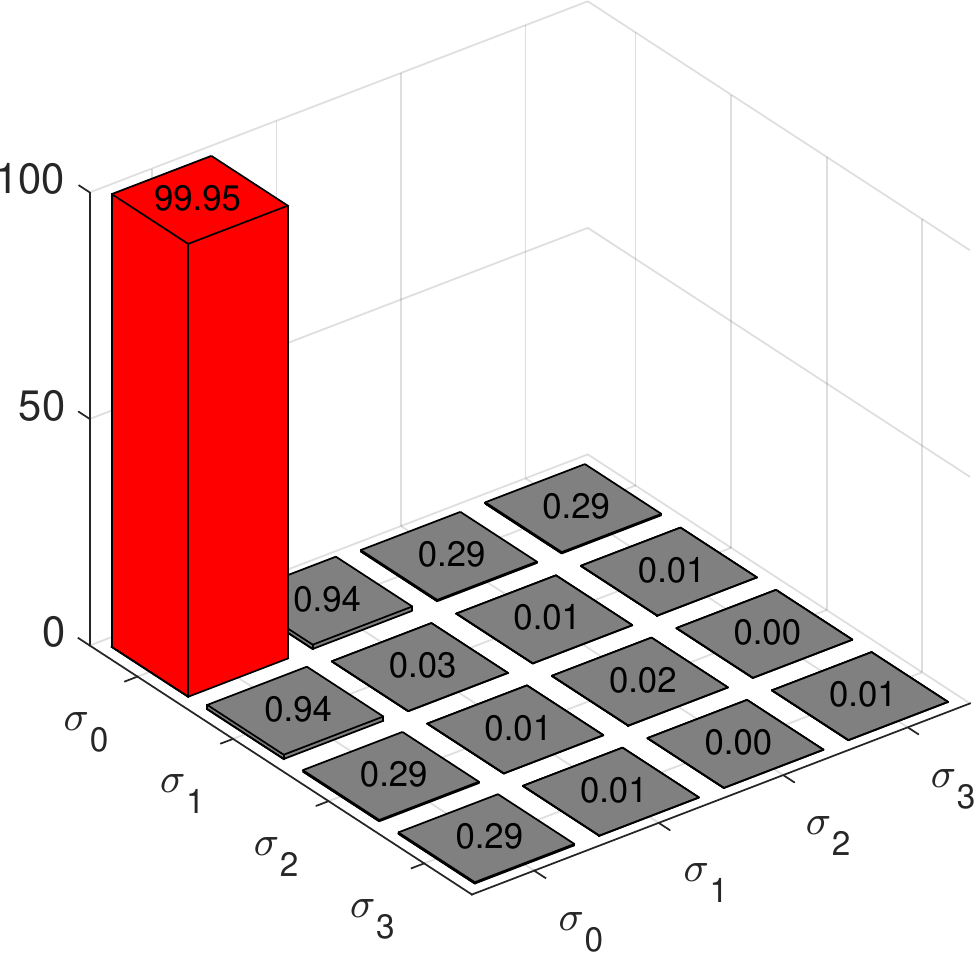}
    \caption{\textbf{Process matrix for the conversion setup.} The absolute values of the individual components are shown. The corresponding process fidelity is 99.947(2)\%.}
    \label{fig:Prozessmatrix}
\end{figure}

\subsubsection*{Conversion Induced Background}
For our DFG-process, the dominant noise source is Anti-Stokes Raman noise \cite{Zaske2011S, Krutyanskiy2017S, Kuo2018S}. To reduce the noise, we chose a waveguide with a comparably low phase-matching temperature of 19°C and the narrowband filtering stage.
For quantifying the conversion induced noise, the 854\,nm input light was blocked. The remaining count rate was integrated over 15 minutes for different pump powers. The dark-count corrected count rates measured on both SNSPD channels are shown in Supplementary Figure \ref{fig:Noiserate}. The actually generated noise rate is inferred from these numbers by dividing by the detection efficiencies of the detectors (31\%/35\%, here the detection efficiency was set to a different value than in the main experiment), the transmission through the waveplates and the Wollaston prism (98\%) and the coupling efficiency to the detector fibers (90\%). The larger error bars for the total noise rate stem from the uncertainty in the detection efficiency of 3\%. The theoretical curve assumes backconversion of 1550\,nm noise photons to 854\,nm which then get blocked by the filtering stage \cite{Maring2018S}. Thus, the curve does not show a linear behaviour as expected for ASR-noise but flattens at higher pump powers which matches well with the experimental data. The resulting noise rate at the working point of 1.1\,W is 24(4)\,photons/s.
\begin{figure}
    \centering
    \includegraphics[width=0.45\textwidth]{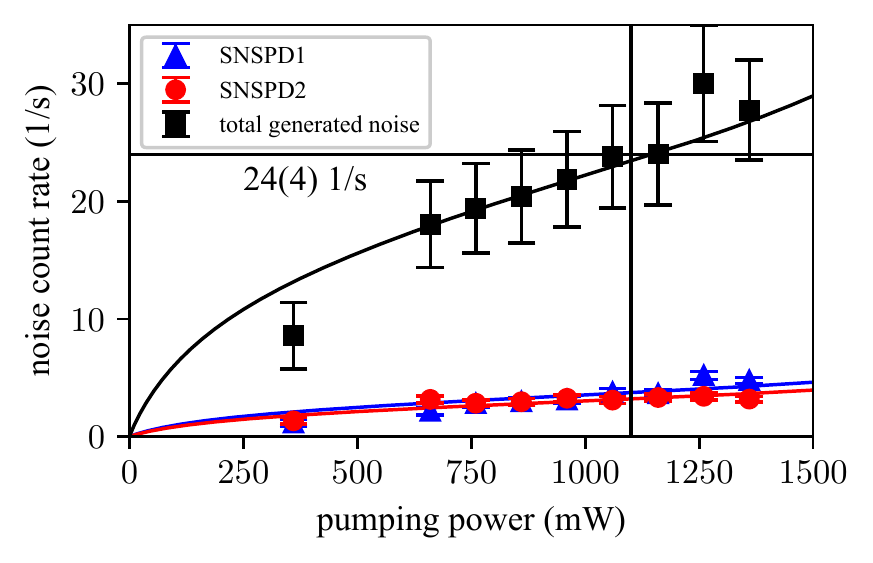}
    \caption{\textbf{Conversion induced noise rate.} Shown are the measured dark-count subtracted count rates for SNSPD channel 1 (blue) and channel 2 (red) as well as the total generated noise-rate. }
    \label{fig:Noiserate}
\end{figure}

\subsection*{Supplementary Notes 1: Setup Losses}

\begin{figure}[h]
    \centering
    \includegraphics[width=0.45\textwidth]{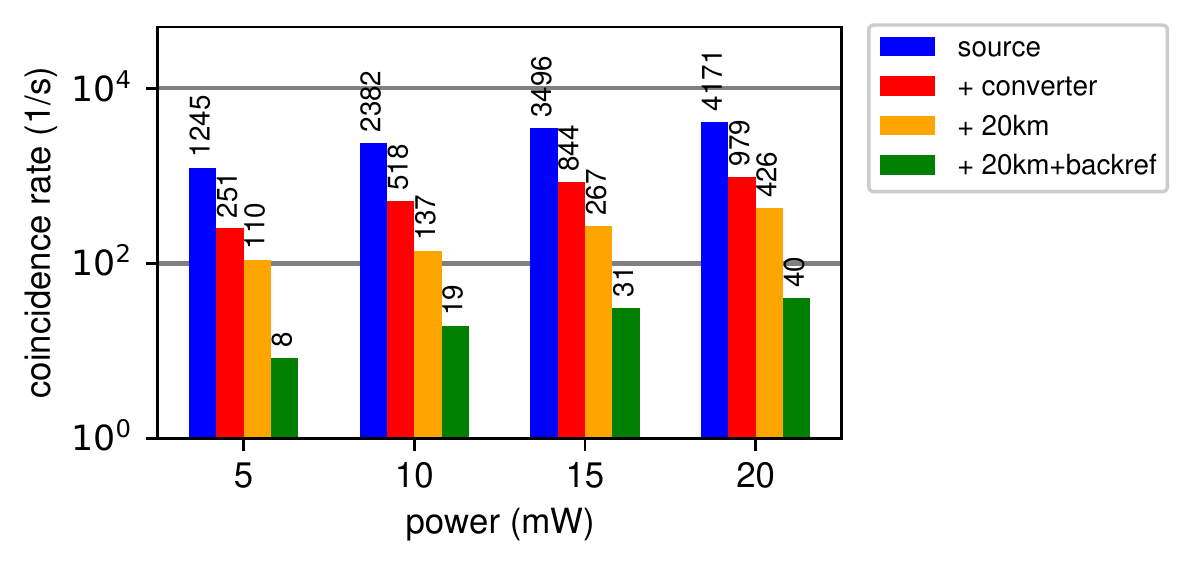}
    \caption{\textbf{Measured coincidence rate}. Comparison of the measured coincidence rate for the four different measurement setups taking $1.5\tau$ of the photon wave packet.}
    \label{fig:coincidence rate}
\end{figure}

\begin{table}[h]
\caption{Transmission and Efficiencies of individual components in setup d)}
\begin{tabular}{c c}	
	\hline
	\hline
	Fiber-fiber couplings & 76\% \\ 
	Fiber-BS (2x) & 50\% \\
	Transmission 854nm fiber (2x) & 87\% \\
	Conversion (2x) & 60\% \\
	Transmission 20km fiber (2x)& 42\% \\
	Retroreflector & 87\%\\
	\hline
	Overall: source output to APD & 0.8\% 
	\end{tabular}
	\label{tab:Efficiencies}
\end{table} 

In Supplementary Figure \ref{fig:coincidence rate} the measured coincidence rates for the four configurations are shown. From that, we estimate the total setup efficiencies. The individually measured component transmissions and efficiencies are shown in Supplementary Table \ref{tab:Efficiencies} for case 4, the back-conversion setup. Fiber to fiber couplings in the source lab account for 76\% transmission. The fiber beam splitter separating the back-converted photons have a transmission of 25\% as the photons pass it twice. The fiber connecting source and converter lab has a transmission of 87\% including coupling efficiencies. As the filtering stage is not used, the conversion efficiency is at 60\%—the 20\,km of fiber and the retroreflector account for 42\% and 87\%, respectively. In total, that leads to an overall setup transmission of 0.8\%. The differences in the setup transmissions inferred from Supplementary Figure \ref{fig:coincidence rate} can be explained by the drifting of the pump laser of the photon pair source.

\subsection*{Supplementary Notes 2: Density matrices}
In Supplementary Figure \ref{fig:density_matrices}, reconstructed density matrices for the different setups are shown exemplary.
\begin{figure*}
    \centering
    \begin{subfigure}{0.85\textwidth}
        \includegraphics[width=\textwidth]{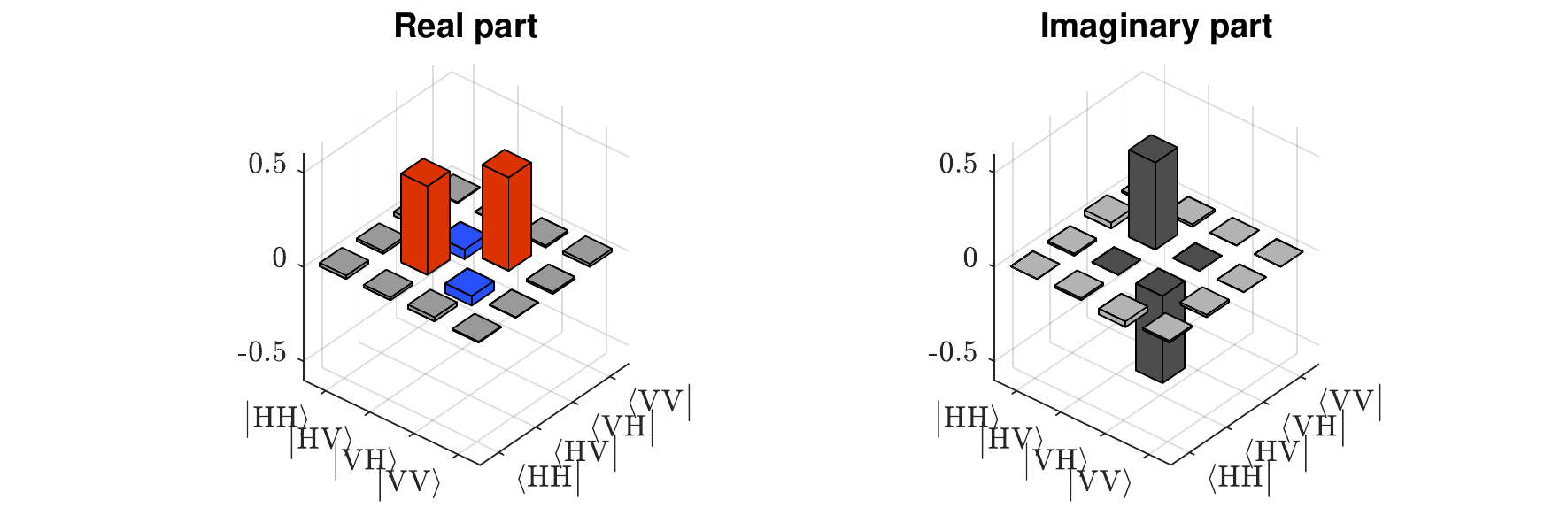}
        \subcaption{source}
    \end{subfigure}
    
    \begin{subfigure}{0.85\textwidth}
        \includegraphics[width=\textwidth]{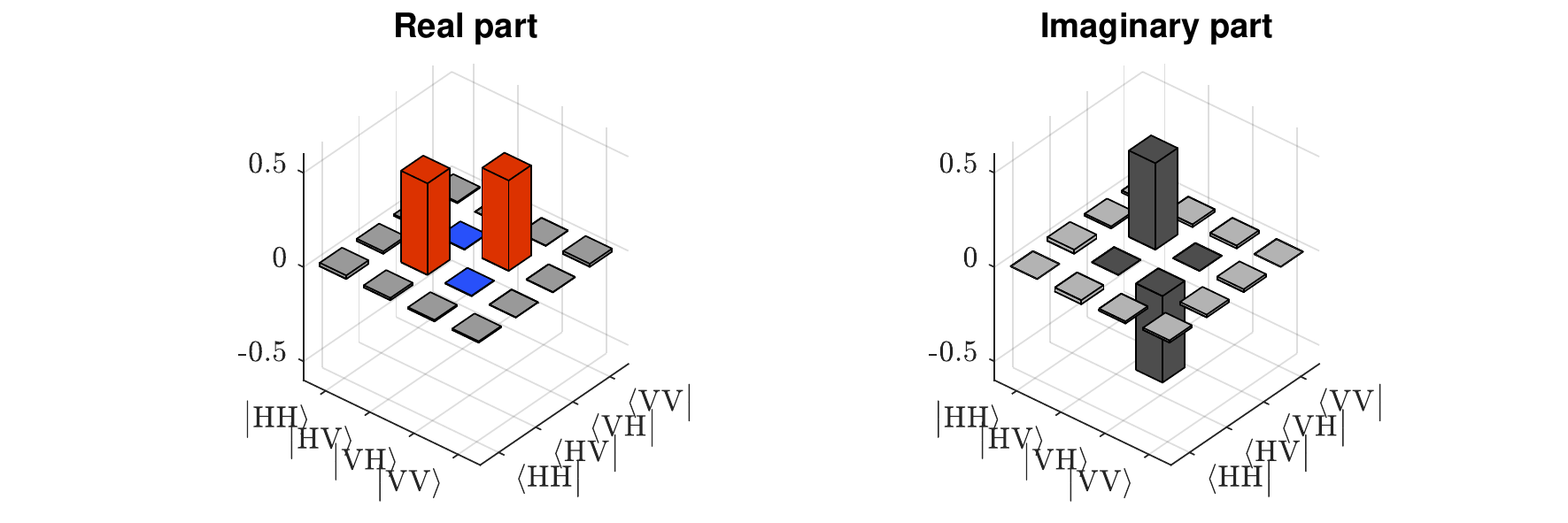}
        \subcaption{source + converter}
    \end{subfigure}\\
    \begin{subfigure}{0.85\textwidth}
        \includegraphics[width=\textwidth]{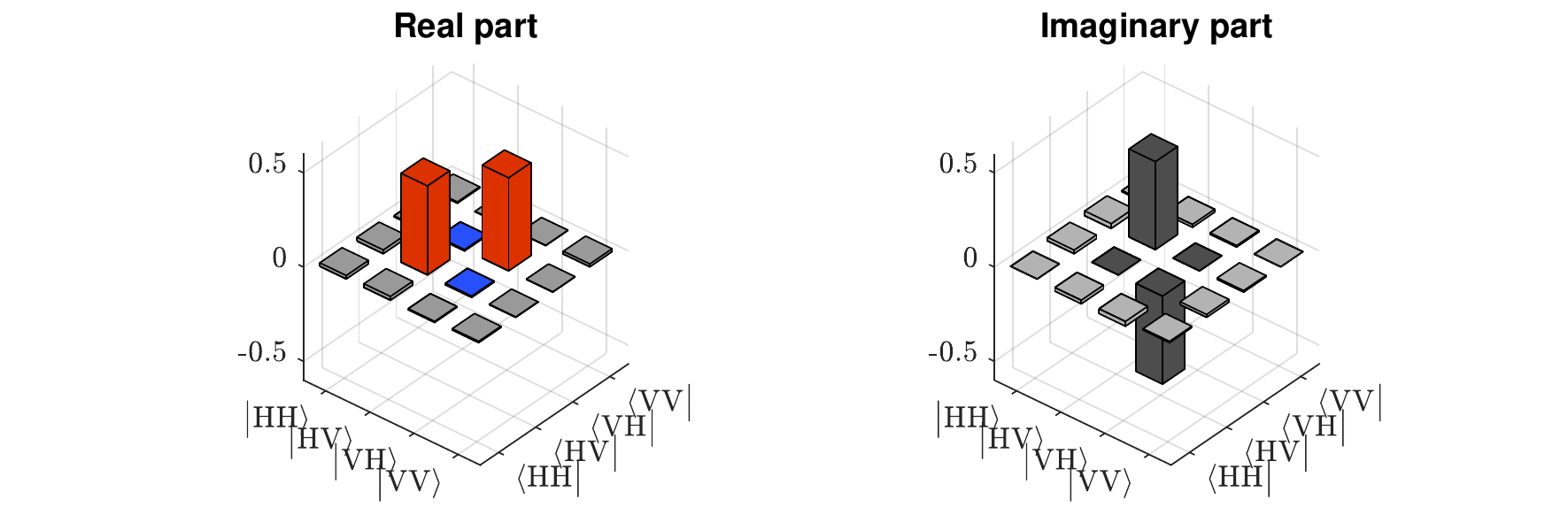}
        \subcaption{source + converter + 20km}
    \end{subfigure}\\
    \begin{subfigure}{0.85\textwidth}
        \includegraphics[width=\textwidth]{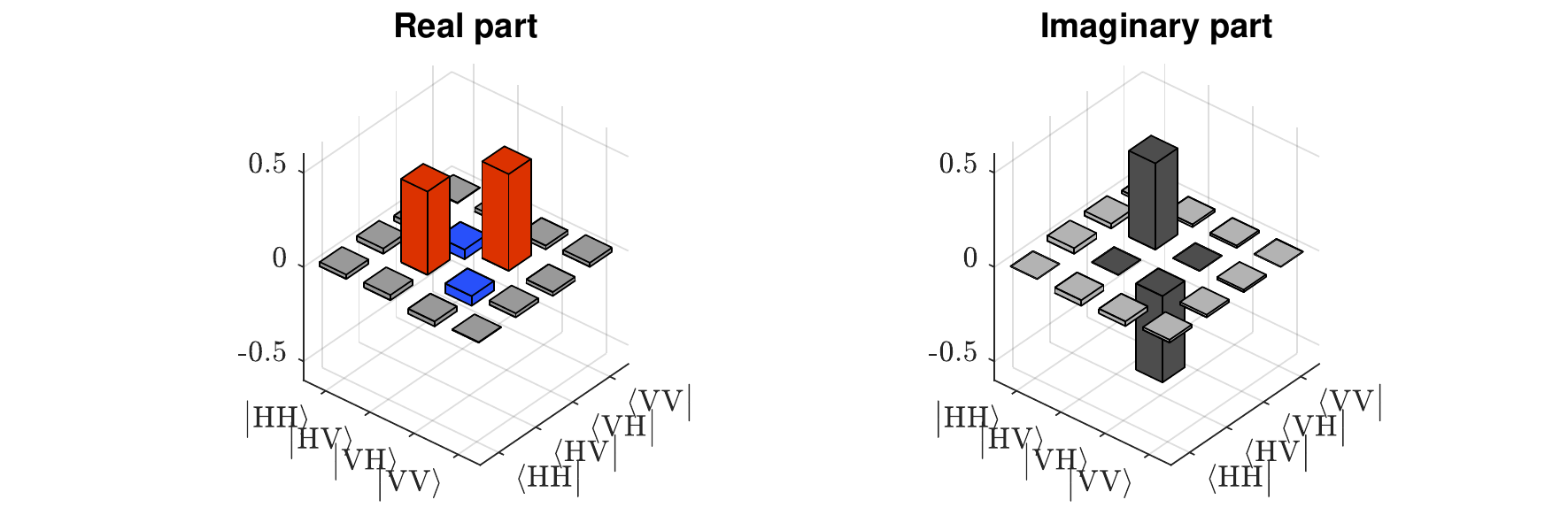}
        \subcaption{source + converter + back reflection + 40km}
    \end{subfigure}
    
    \caption{\textbf{Density matrices for the four different setups.} The shown density matrices corresponds to a time window $\Delta t = 5\tau_\text{mean}$ and a pump power of 20 mW.}
    \label{fig:density_matrices}
\end{figure*}

\clearpage
\subsection*{Supplementary References}